# Time interval measurements by uniformly accelerating observers (non-longitudinal case)

Stefan Popescu[1] and Bernhard Rothenstein[2]

1) Siemens AG, Erlangen, Germany
2) Politehnica University of Timisoara, Physics Department, Timisoara, Romania

***Abstract.*** *We consider two non-longitudinal Doppler Effect experiments. The first one involves a stationary source of monochromatic light located at the origin O of the K(XOY) inertial reference frame and an observer R who performs the hyperbolic motion at a constant altitude. The second involves a stationary observer located at the origin O and a monochromatic source of light that performs the hyperbolic motion. In both cases we compare the relationship between emission and reception times of the same light signal and the relationship between the emission and the reception time intervals of two successive light signals.*

## 1. Introduction

In a previous paper [1] we have studied the following three scenarios: stationary receiver and accelerating source of light, stationary source of light and accelerating receiver and stationary source of acoustic waves and accelerating observer. In all cases the motion is the well known hyperbolic one [2] that takes place along the line that joins source and receiver. The purpose of the present paper is to extend the problem to the non-longitudinal case when the velocity and the instantaneous position vector make a time dependent angle. The problem was studied by Neutze and Moreau [3] who take into account the fact that a moving observer is not able to measure the time interval between the reception of two successive wave-crests from the same point in space (not instantaneous measurement). They consider only the case when the moving observer receives the first wave-crest at a zero time when his speed is equal to zero, receiving the second one being in motion. The purpose of our paper is to generalize the results of Neutze and Moreau [3] considering all the successive time intervals measured by the moving observer.

We propose the following scenarios:
1. Stationary source of light S located at the origin O of the inertial reference frame K(XOY) and an accelerating receiver moving at constant altitude $y = h$.

2. Accelerating source of light moving at constant altitude $y = h$ and a stationary receiver located at the origin O of the K(XOY) inertial reference frame.

In these scenarios the accelerating motion along the OX axis is described by the following equations:

$$x = \frac{c^2}{g}\left(\cosh\frac{gt'}{c} - 1\right) \quad \text{respectively} \quad x = cT\left(\cosh\frac{t'}{T} - 1\right) \tag{1}$$

$$t = \frac{c}{g}\sinh\frac{g}{c}t' \quad \text{respectively} \quad t = T\sinh\frac{t'}{T} \tag{2}$$

where $g$ represents the constant acceleration of the moving object (source or receiver) with respect to its instantaneously commoving inertial frame, $t$ and $t'$ are the readings of a stationary clock C and respectively of a clock C′ commoving with the accelerating object when they are located at the same point in space. In the right side equations we use the notation $T = \frac{c}{g}$ in order to simplify the equations and to better underline the physical significance of various terms. Here $T$ may be interpreted as a "magic" time interval required accelerating an object with constant acceleration $g$ until it reaches the speed $c = gT$ in a non-relativistic approach.

When represented in world coordinates [$t$, $x/c$] the equations (1) and (2) correspond to the parametric equations of a conjugate hyperbola. Accordingly this motion is also known as the "hyperbolic motion". It begins at $x = +\infty$, $t = -\infty$ with $V = c$ when the decelerating object approaches the origin with proper acceleration $-g$ until it reaches the rest at $t = 0$ and $x = 0$. Thereafter for $t > 0$ the moving object reverse direction receding the origin with proper acceleration $+g$ and reaching $V = c$ as $x \rightarrow +\infty$, $t \rightarrow +\infty$ in accordance with the requirements of special relativity.

The invariance of distances measured perpendicular to the direction of relative motion requires that

$$y = y'. \tag{3}$$

**2. Stationary source of light and accelerating receiver**

For keeping the equations as simple as possible we consider that the light signal emitted by the source S at $t = t_e$ will be received at $t = t_r$ by the observer located at the point R($x_r$,$h$).

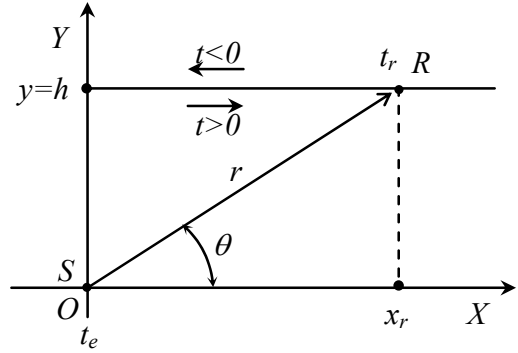

**Figure 1**. The stationary source of light located at the origin O and an accelerating receiver maintaining constant altitude *y=h*

Using the notations introduced above we obtain the world coordinates of the reception point as:

$$x_r = cT\left(\cosh\frac{t'_r}{T} - 1\right) \quad (4)$$

$$t_r = T \sinh\frac{t'_r}{T}. \quad (5)$$

Furthermore Pythagoras' theorem applied to Figure 1 leads to

$$x_r^2 + h^2 = c^2(t_r - t_e)^2. \quad (6)$$

Taking into account (4) and (5) and solving equation (6) for $t'_r$ we find the following expression for the reception time in frame K':

$$t'_{r\pm} = T \cdot \ln\frac{2T^2 + T_h^2 - t_e^2 \pm \sqrt{4T_h^2 T^2 + (T_h^2 - t_e^2)^2}}{2T(T - t_e)} \quad (7)$$

Here $T_h = \frac{h}{c}$ is the time interval required for the circular wave-front of a light signal to reach for the first time the altitude *h* of the receiver trajectory. With (7) we have two mathematical solutions for the reception time, depending on the sign selected before the square root. We represent in Figure 2 the two mathematical solutions of (6) for $T = T_h = 1$ and for different values of the emission time.

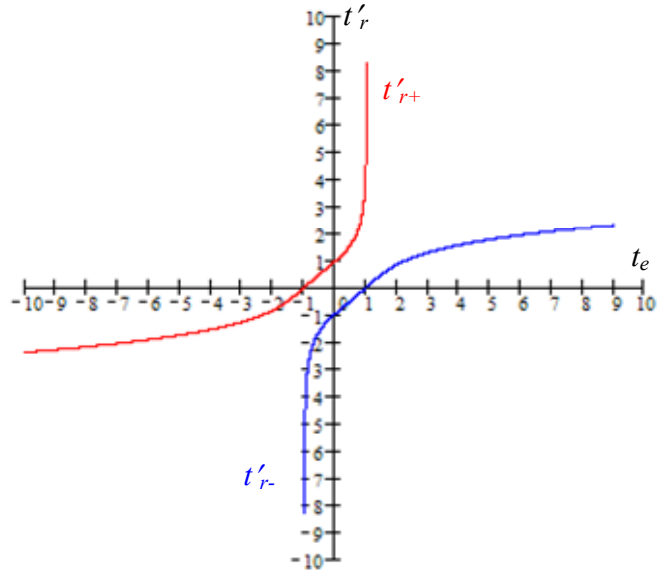

**Figure 2**. The two mathematical solutions for the reception time depending on the sign in front of the square root

Subsequently we investigate which solution is physically valid under different circumstances. We will solve this problem in a world coordinate diagram. For this we calculate and represent the world line of the virtual point $W(x_w,t)$ representing the intersection between the circular wave-front of the propagating light signal and the receiver trajectory.

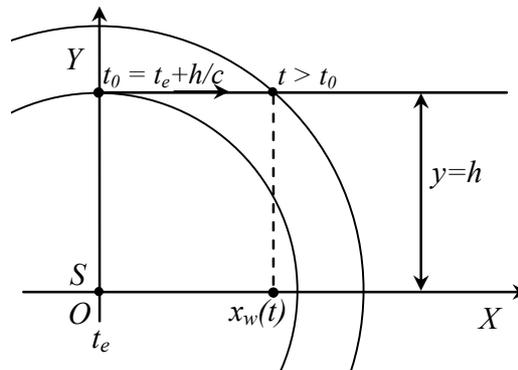

**Figure 3**. The intersection between the circular wave-front of the light signal and the trajectory of the accelerating receiver at altitude $y=h$

Pythagoras' theorem applied to Figure 3 leads to
$$x_w^2 + h^2 = c^2(t-t_e)^2. \tag{8}$$
which solved for $x_w$ returns the horizontal position of the $W$ point as:
$$x_w(t) = \sqrt{c^2(t-t_e)^2 - h^2} \tag{9}$$

and its speed:

$$v_w(t) = \frac{c}{\sqrt{1 - \frac{h^2}{c^2(t-t_e)^2}}} \qquad (10)$$

Equation (8) describes a hyperbola while equation (10) reveals a supra-luminal speed for the virtual point of intersection. This result doesn't disregard special relativity because no transport of energy (information) takes place in the direction in which the observer moves. We illustrate both functions in Figure 4.

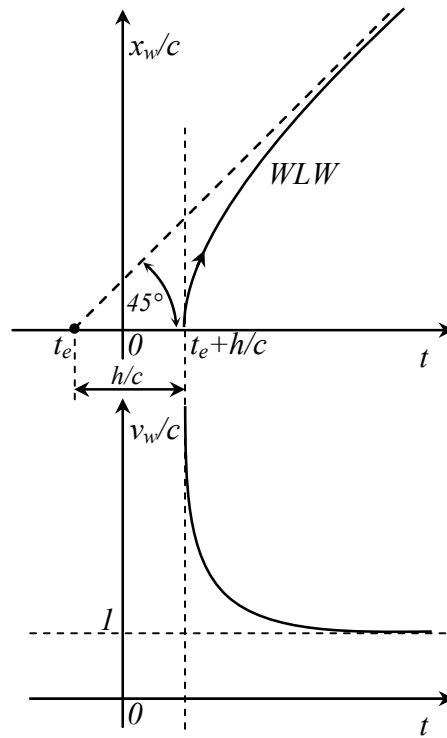

**Figure 4**. The world line and the speed of the intersection point between the wave-front of the light signal and the trajectory of the accelerating receiver

The resulted supra-luminal speed of the wave-front point *W* suggests that we may have at most one superposition between this point and the accelerated receiver, therefore at most one reception time and not two as mathematically given by (7). Further conclusions result after depicting the world line of the accelerating receiver (*WLR*) and the word line of the *W* point (*WLW*) in a common world diagram.

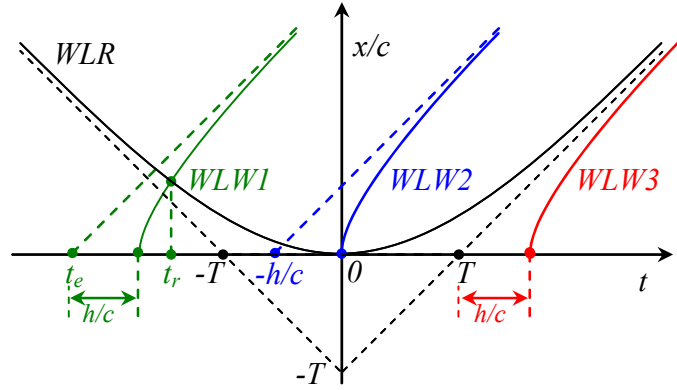
**Figure 5**. The world diagram for different emission times

Figure 5 depicts three emission-receptions events. If $t_e \leq T$ then we have an unique reception time, which is 0 for $t_e = -h/c$ and $+\infty$ for $t_e = T$. For $t_e > T$ the communication between source and receiver isn't possible anymore. It is remarkable to notice here that the last emission time for which a reception is still possible doesn't depend on the receiver altitude but it depends only on receiver acceleration $g$ and the speed of light, having the same value as for $h = 0$, which is the longitudinal Doppler scenario that we analysed in [1].

Now that we identified the physically valid solution of equation (6) which gives the reception time $t'_r$ as:

$$t'_r = T \cdot \ln \frac{2T^2 + T_h^2 - t_e^2 + \sqrt{4T_h^2 T^2 + (T_h^2 - t_e^2)^2}}{2T(T - t_e)} \tag{7'}$$

we depict in figure 6 the variation of reception time $t'_r$ with the emission time $t_e$ for different values of $T_h = h/c$ and for $T = 1$.

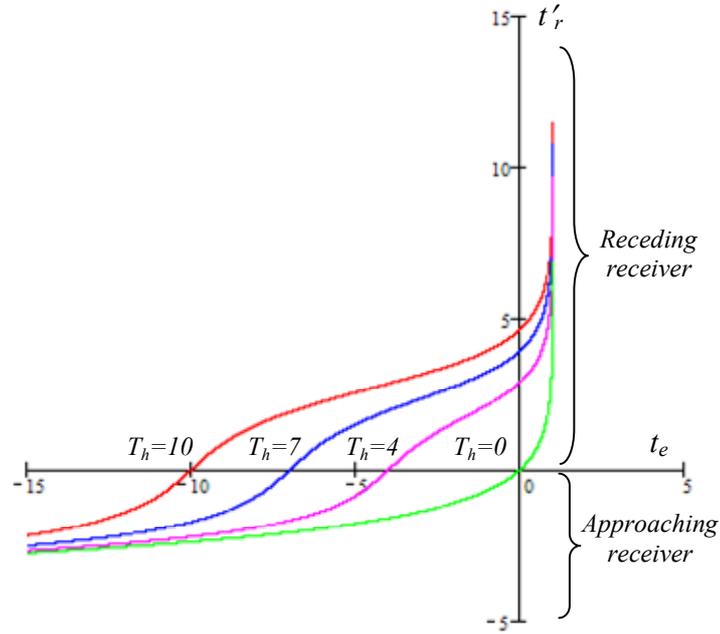

**Figure 6**. The variation of reception time with the emission time for $T = 1$ and for different values of $T_h$ mentioned on the corresponding curves

Supposing that the stationary source emits periodic light signals at time instants $t_{e,N} = N \cdot T_e$ then the accelerating observer receives the $N^{th}$ light signal when his wrist watch reads $t'_{r,N}$. Using (7') we calculate the successive reception instants measured by R as well as the time interval between them. This allows us to introduce the concept of Doppler factor defined as the quotient between the proper reception and the proper emission periods as:

$$D(N) = \frac{T'_r(N)}{T_e} = \frac{t'_{r,N} - t'_{r,N-1}}{T_e} \tag{11}$$

Figure 7 shows the variation of $D$ with the order number $N$ and different values of $T_e$. Figure 8 illustrates the variation of the Doppler factor $D$ with the emission order $N$ and the influence of $T_h = h/c$.

Figure 7 and 8 show that for a constant emission period the received period increases until the receiver is at rest (local maximum), but thereafter we have a time window in which the received period decreases before increasing again (local minimum). For negative emission times we have $D < 1$ i.e. the proper reception period is lower than the proper emission period, whilst for positive emission times we have $D > 1$, thus the proper reception period is larger than the proper emission period. It is also amazing to notice in figure 7 that we have a region of inverted monotony wherein for the same order number $N$ the reception period becomes longer as the emission period goes shorter.

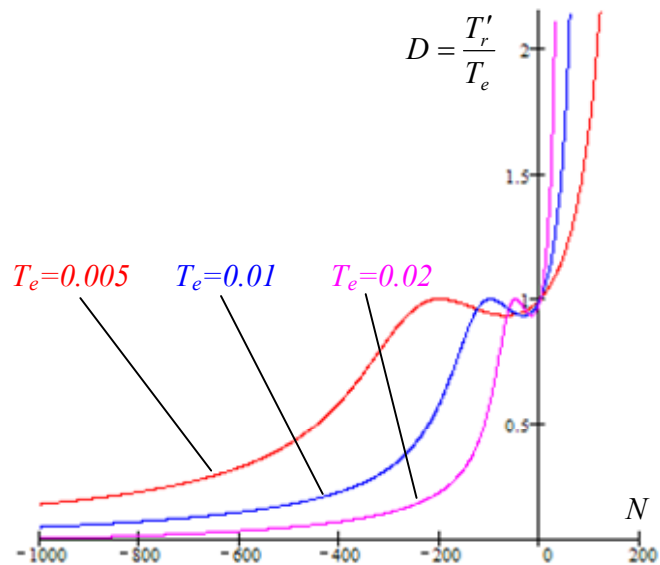

**Figure 7**. The variation of Doppler factor $D$ with the emission order $N$ for $T_h = T = 1$ and for different values of $T_e$

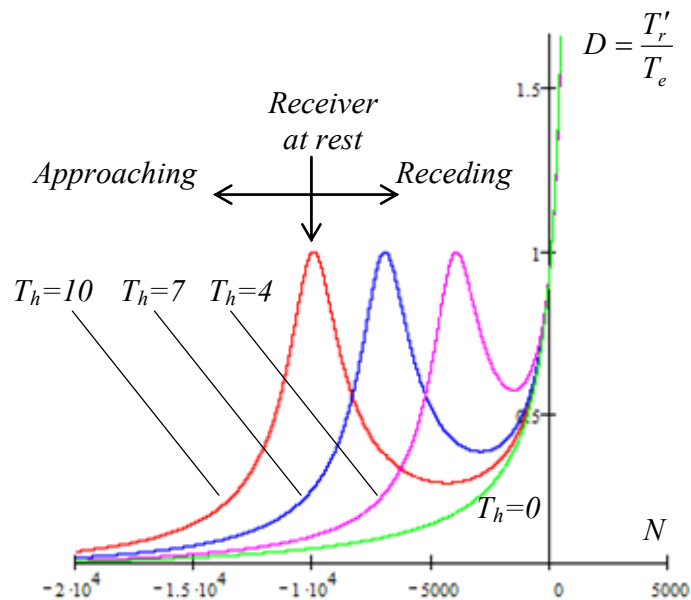

**Figure 8**. The variation of Doppler factor $D$ with the emission order $N$ for $T = 1$, $T_e = 10^{-3}$ and different values of $T_h = h/c$

We notice in figure 8 that for very early emission times (when the approaching receiver is far away from source) or just before the interruption of communication with the receding receiver (when $t_e \rightarrow T$) the influence of *h/c* diminishes and all curves approach the one-dimensional case with *h* = 0 (longitudinal Doppler Effect). In-between these regions *h/c* achieves a stronger influence with the receive period reaching a local maximum whose peak value is independent on the receiver altitude *h* being equal to the emission period $T_e$. This maximum occurs in a region where the relativistic effects may be ignored, more exactly when the receiver is at rest, which means that $N_{peak} = -\frac{T_h}{T_e} = -\frac{h}{cT_e}$.

Furthermore we notice in figures 7 and 8 a further interesting aspect of this scenario, namely that for *h* ≠ 0 we have two situations when the reception period $T'_r$ equals the emission period $T_e$. The first one occurs with the observer at rest receiving the signals emitted at $t_e = -T_h = -h/c$ and the second one occurs with a receding observer receiving the signals emitted at $t_e = 0$ when the observer was at rest. More surprising in between these two time instances we notice a region in which *D* < 1. Thus, contrary to longitudinal Doppler, the proper receive period may be shorter than the proper emission period also during the receding phase.

### 3. Stationary receiver and accelerating source of light

For keeping the equations as simple as possible we consider that the light signal emitted by the accelerating source S at $t = t_e$ and at the point S($x_e$,*h*) will be received at time $t = t_r$ by the stationary observer located at the origin O.

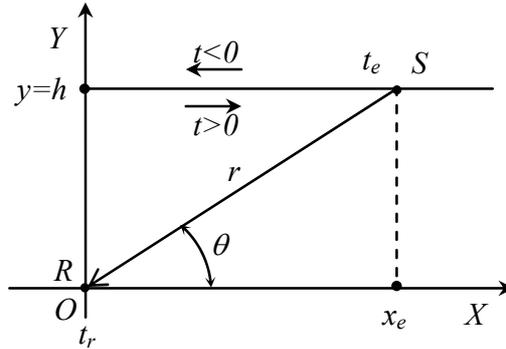

**Figure 9**. The stationary receiver located at the origin O and an accelerating source of light maintaining constant altitude *y=h*

Using the notations introduced above we obtain the world coordinates of the emission point as:

$$x_e = cT\left(\cosh\frac{t'_e}{T} - 1\right) \qquad (14)$$

$$t_e = T \sinh \frac{t'_e}{T}. \tag{15}$$

Furthermore Pythagoras' theorem applied to Figure 9 leads to
$$x_e^2 + h^2 = c^2(t_r - t_e)^2. \tag{16}$$

Taking into account (14) and (15) and solving equation (16) for $t_r$ we find the following expression for the reception time in frame K:

$$t_{r\pm} = T \cdot \left( \sinh \frac{t'_e}{T} \pm \sqrt{\left(\cosh \frac{t'_e}{T} - 1\right)^2 + \left(\frac{T_h}{T}\right)^2} \right) \tag{17}$$

Here $T_h = \frac{h}{c}$ is the time interval required for the circular wave-front of a light signal to reach for the first time the OX axis. With (17) we have again two mathematical solutions for the reception time. We represent in Figure 10 the two mathematical solutions of (16) for $T = T_h = 1$ and for different values of the emission time.

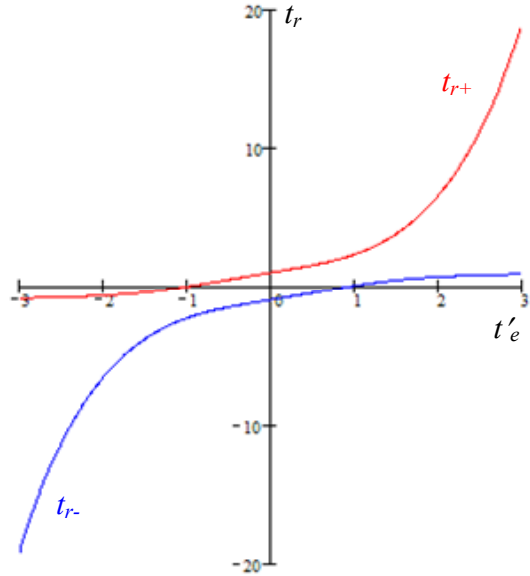

**Figure 11**. The two mathematical solutions for the reception time

We apply the same method as for the first scenario in order to investigate for the physically valid solutions and we obtain the world diagram illustrated in Figure 12. It shows that for each emission time we have a unique reception time $t_r \geq -T$. Furthermore for $t'_e = 0$ we have $t_r = T_h = h/c$. Therefore the single physically valid solution for (16) is :

$$t_r = T \cdot \left( \sinh \frac{t'_e}{T} + \sqrt{\left( \cosh \frac{t'_e}{T} - 1 \right)^2 + \left( \frac{T_h}{T} \right)^2} \right) \tag{17'}$$

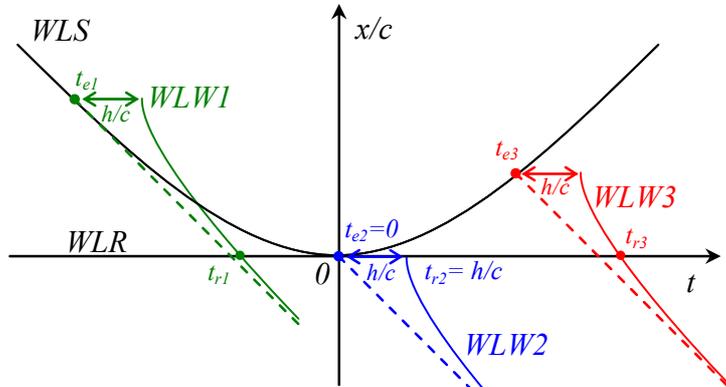

**Figure 12**. The world diagram for different emission times

We depict in figure 13 the variation of reception time $t_r$ with the emission time $t'_e$ for different values of $T_h = h/c$.

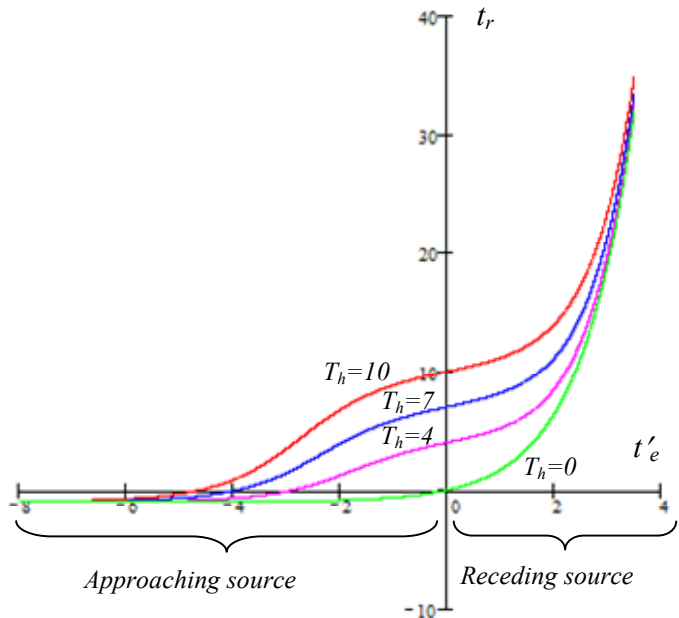

**Figure 13**. The variation of reception time with the emission time for $T = 1$

We notice in figure 13 that for very early or very late emission times (when the source is far away from receiver) the influence of $h/c$ diminishes and all curves approach the one-dimensional case with $h = 0$ (longitudinal Doppler Effect). If the distance source-receiver is relatively small then $h/c$ achieves more significance.

Supposing now that the accelerating source emits periodic light signals at periodic time instants $t'_{e,N} = N \cdot T'_e$ then the stationary observer receives the $N^{th}$ light signal when his wrist watch reads $t_{r,N}$. Using (17') we calculate the successive reception instants measured by R as well as the time interval between them. This allows us to calculate the Doppler factor $D$ as:

$$D(N) = \frac{T_r(N)}{T'_e} = \frac{t_{r,N} - t_{r,N-1}}{T'_e} \tag{18}$$

and Figure 14 shows its variation with the order number $N$ for different values of $T'_e$. Figure 15 shows the influence of $T_h = h/c$.

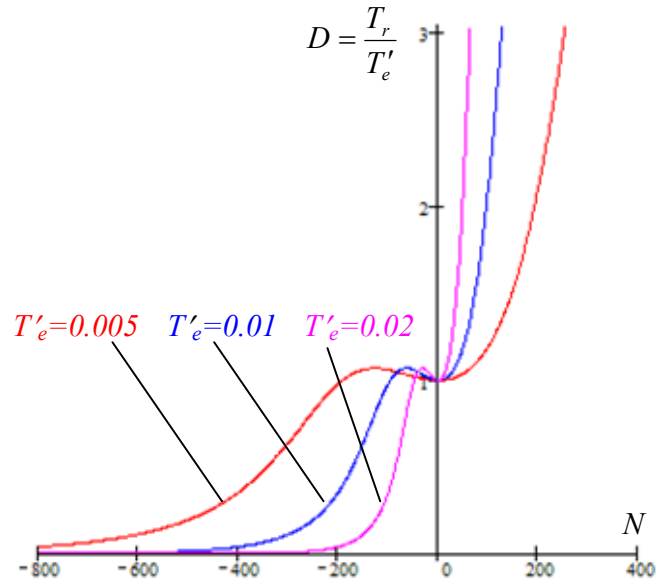

**Figure 14**. The variation of Doppler factor with the emission order $N$ for $T_h = T = 1$ and for different values of $T'_e$

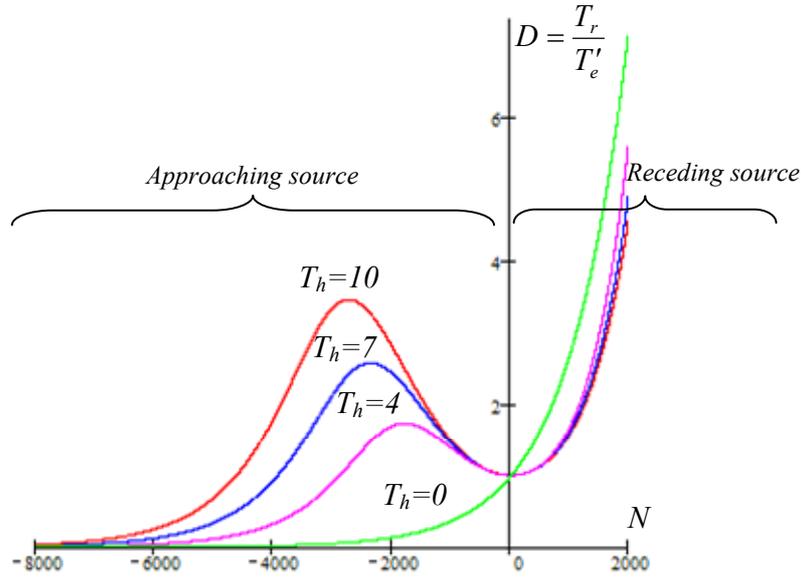

**Figure 15**. The variation of Doppler factor $D$ with the emission order $N$ for $T = 1$ and $T'_e = 10^{-3}$ and different values of $T_h = h/c$

In figure 14 and 15 we observe the same oscillatory behaviour of the reception period, occurring for $h \neq 0$ and creating two points when the proper reception period $T_r$ equals the proper emission period $T'_e$. The first one occurs with the source at rest separating the approaching and receding phases and the second one occurs with the source approaching the rest position. These points delimit a local maximum region. Moreover, we find that in this region we have $D > 1$. Thus, contrary to longitudinal Doppler effect, the proper receive period may be longer than the proper emission period also during the approaching phase.

## 6. Conclusions

We reveal some new and interesting aspects related to the behaviour of the time interval between the reception of two successive light signals as measured by an accelerating observer when emitted by a stationary source of light and of the time interval between two successive light signals as measured by a stationary observer when emitted by an accelerating source of light. In both cases the accelerated motion is the hyperbolic one and it takes place at a constant altitude. Our approach teaches how to put in equations a given physical scenario. We also present an innovative method allowing to choose the valid physical solution among multiple mathematically solutions derived by formulas. In what is concerning the range of contact between source and receiver we found striking similarities between the scenarios we analyse here and the longitudinal Doppler effect we analysed earlier [1].